\documentclass[conference]{IEEEtran}
\IEEEoverridecommandlockouts

\ifCLASSINFOpdf
\else
\fi

\usepackage{amsmath}
\usepackage{amssymb}
\usepackage{multirow} 

\usepackage[usenames,dvipsnames]{xcolor}
\definecolor{CellOrange}{HTML}{F58518}

\usepackage{cite}
\usepackage{tikz}
\usepackage{quantikz}
\usepackage[table]{xcolor}
\usepackage{diagbox}
\usetikzlibrary{positioning, arrows.meta}

\begin{document}

\title{Balancing Expressivity and Overfitting in Quantum Gaussian Process Regression} 

\author{\IEEEauthorblockN{Saasha Joshi} 
\IEEEauthorblockA{\textit{CMC Microsystems} \\
Toronto, ON, Canada \\
saasha.joshi@cmc.ca} 
\and
\IEEEauthorblockN{Udson C. Mendes}
\IEEEauthorblockA{\textit{CMC Microsystems} \\
Sherbrooke, QC, Canada \\
udson.mendes@cmc.ca}
\and
\IEEEauthorblockN{Luke C. G. Govia}
\IEEEauthorblockA{\textit{CMC Microsystems} \\
Waterloo, ON, Canada \\
luke.govia@cmc.ca}
}

\maketitle

\begin{abstract}
Active learning is a paradigm of machine learning that can be utilized to model expensive black-box functions by training a surrogate model from actively queried training points. The performance of this framework depends heavily on the choice of the surrogate model. When the surrogate is Gaussian Process Regression (GPR), its performance is largely determined by the expressivity of the underlying kernel. In this work, we investigate the peculiarities of using a quantum kernel to change the computational dynamics of active learning with GPR, focusing on the sensitivity to regularization by hyperparameter tuning. While generic, unstructured kernels suffer from exponential concentration at large scale, we empirically demonstrate that even at small scale overfitting can collapse GPR performance. Kernel regularization can be used to counteract this effect, but due to the smoothness of the quantum fidelity kernel, regularization must be carefully chosen to balance expressivity and overfitting. Our qualitative results transfer to the practical application of restricted quantum kernels designed to avoid exponential concentration and also present the kinds of noise that may be valuable to kernel training in near-term quantum devices.
\end{abstract}


\IEEEpeerreviewmaketitle

\section{Introduction}

Many problems in science and engineering are formulated as black-box functions that lack an explicit mathematical expression~\cite{garnett2023bayesian}, with computationally expensive evaluation, such that their optimization is constrained by a limited evaluation budget~\cite{grill2015black}. Such black-box functions arise in a wide range of applications, including simulations in engineering design, drug discovery, financial modeling, and machine learning~\cite{feurer2019hyperparameter}.
Surrogate models, computationally inexpensive approximations of black-box functions, are often used in place of direct black-box function evaluation, but these are limited by two competing challenges: $(1)$ accurately capturing the underlying structure of the unknown function and $(2)$ minimizing the number of expensive black-box function evaluations used to build the surrogate model. 

Active learning is a paradigm of machine learning utilized to build surrogate models by actively querying sequential training points guided by uncertainty predictions from the surrogate model~\cite{schulz2018tutorial}. A common choice for the surrogate model is Gaussian Process Regression (GPR)~\cite{schulz2018tutorial, williams2006gaussian, gramacy2020surrogates}, a non-parametric regression model that provides the required predictive estimate and uncertainty quantification through the mean and variance of a multi-variate Gaussian distribution describing the unknown function output. 
The performance of GPR depends critically on the choice of the covariance function, or kernel~\cite{gramacy2020surrogates, williams2006gaussian}, which describes the similarity between data inputs and encodes correlation patterns of the black-box function. Kernel hyperparameters also play an important role and poor choice can lead to underfitting or overfitting of training data, thus affecting the predictive accuracy of the surrogate~\cite{williams2006gaussian}. 

Some approaches to Quantum Machine Learning (QML) use quantum kernels~\cite{havlivcek2019supervised, schuld2019quantum}, which encode input data into the high-dimensional Hilbert space of a quantum system. Similarity measures can be defined in a few ways~\cite{balcan2006theory}, but here we focus on using the natural inner product between vectors in Hilbert space. Although quantum kernels have been extensively studied for regression and classification tasks~\cite{havlivcek2019supervised, blank2020quantum}, their integration into the active learning workflow remains relatively underexplored. Rapp and Roth~\cite{rapp2024quantum} explore Quantum Bayesian Optimization (QBO) using a GPR surrogate with a quantum kernel constructed from the Chebyshev feature map~\cite{rapp2024quantum,kyriienko2021solving}. J\"{a}ger et al.~\cite{jager2026provable} leverage symmetry in free-fermionic evolutions to construct a quantum kernel for GPR. While this technique addresses the key issue of exponential concentration, it results in classically tractable kernels.

Broadly speaking, the symptom of exponential concentration, in the absence of noise, is a diagonal kernel that amounts to a model that has overfit the training data. Generally impacting high-dimensional quantum kernels, symmetry restrictions~\cite{henderson2025quantum} aim to alleviate this issue by constraining the size of the active Hilbert space. However, even low-dimensional kernels can become diagonal and effectively concentrate when the input data are simple compared to the expressive power of the kernel. In classical machine learning, this issue is commonly resolved by regularization~\cite{williams2006gaussian, gramacy2020surrogates} to reduce overfitting of the training data.

In this work, we utilize an active learning workflow with quantum GPR (QGPR) to build a surrogate model for a one-dimensional black-box function and study the impact of regularization on surrogate quality. We compare multiple quantum kernels and explore how the qubit count and circuit connectivity impact active learning performance. Our empirical study of low-dimensional kernels is meant to highlight the potential use of regularization to improve performance at \emph{practical} kernel dimensions, but is not itself an exploration of QGPR with high-dimensional kernels in the quantum advantage regime. As regularization involves adding synthetic noise, our work also provides insight into how real noise can be engineered into the quantum kernel and become beneficial in practice.

\section{Problem Formulation}

We consider the problem of modeling a black-box function $f: \chi \subset \mathbb{R}^d \mapsto \mathbb{R}$ in domain $\chi$ for which evaluations are expensive. Let $\mathcal{D}_n = \{(x_i, y_i = f(x_i)\}_{i=1}^n$ denote a set of $n$ input-output pairs obtained from function evaluations. The primary goal is to construct a surrogate model $\hat{f}(x)$ that approximates $f(x)$ across the input domain. To achieve this, the GPR surrogate is used to obtain probabilistic estimates of $f(x)$ in terms of a predictive mean $\mu = \mathbb{E}[\hat{f}(x)]$, where $\mathbb{E}[\cdot]$ is the expectation operator, and variance $\sigma^2$, quantifying the uncertainty of the model prediction.

Since function evaluations are expensive, it is crucial to carefully choose the next evaluation point $x_{n+1}$ used to improve the surrogate model by evaluating $y_{n+1} = f(x_{n+1})$. This is done by optimizing an acquisition function $J(x \mid D_n)$, which guides the selection of the next point to informative regions of the input space. In this work, $J(x \mid D_n)$ selects from the candidate points the one that provides the maximum predictive variance~\cite{schulz2018tutorial}, leading to the exploration of regions with greater uncertainty. The surrogate is subsequently retrained with the updated dataset, and this is repeated until the evaluation budget is exhausted or the model converges.

\section{Gaussian Process Regression}

GPR provides a probabilistic framework for modeling unknown functions and is widely used for active learning. Given a set of input points $X_n = \{x_i\}_{i=1}^{n}$, it is assumed that the function values $Y_n = \{f(x_i)\}_{i=1}^{n}$ follow a multivariate normal distribution $ Y \sim \mathcal{N}(\mu, \Sigma)$~\cite{garnett2023bayesian}. Here $\mu$ is the mean vector with entries $\mu_i = \mu(x_i)$ and $\Sigma_n \in \mathbb{R}^{n\times n}$ is a positive semi-definite covariance matrix defined by the covariance function $\Sigma(\cdot, \cdot)$, where each element is given by $(\Sigma_n)_{ij} = \Sigma(x_i, x_j)$. Given data $\mathcal{D}_n$, the predictive distribution at a new point $x_{n+1}$ is given as $Y(x_{n+1}) \mid \mathcal{D}_n \sim \mathcal{N}(\mu(x_{n+1}), \sigma^2(x_{n+1}))$, with mean and variance given by
\begin{align}
    \mu(x_{n+1}) &= \Sigma(x_{n+1}, X_n) \Sigma^{-1}_n Y_n \label{eq:mean-variance} \\
    \nonumber\sigma^2(x_{n+1}) &= \Sigma(x_{n+1}, x_{n+1}) - \Sigma(x_{n+1}, X_n)\Sigma^{-1}_n \Sigma(X_n, x_{n+1})
\end{align}
Here, $\Sigma(x_{n+1}, X_n)$ is a $1 \times n$ covariance vector between $x_{n+1}$ and $X_n$.

\subsection{Hyperparameters: Signal and Noise Variance}

Although GPR is a non-parametric model, its behavior is influenced by hyperparameters that fine tune the model by regularization. In this work, we focus on two hyperparameters, known as \emph{signal variance} $\tau^2$ and \emph{noise variance} $\lambda^{2}$.

\subsubsection{Signal Variance}
This hyperparameter scales the covariance matrix, controlling the amplitude of the correlations the surrogate represents~\cite{gramacy2020surrogates} so that $Y \sim \mathcal{N}(\mu, \tau^2 \Sigma)$. 
If $\tau^2$ is too small, the model assumes that the underlying function remains close to the mean, preventing it from capturing large variations in the data. Conversely, if $\tau^2$ is too large, the model varies widely and can fit individual data points too aggressively, increasing the chances of overfitting~\cite{williams2006gaussian}. 

\subsubsection{Noise Variance}
This hyperparameter acts as synthetic ``noise'' on the diagonal of the covariance matrix, and improves its numerical stability~\cite{gramacy2020surrogates}. 
If $\lambda^2$ is too small, the model can put too much weight on the observed data and attempt to interpolate this data exactly, which can lead to overfitting. Conversely, if $\lambda^2$ is too large, the model observes data variation as noise, resulting in overly smooth predictions and underfitting. When combined with signal variance, the kernel entries are given by
\begin{equation}
    \Sigma_{ij} = \tau^2\Sigma(x_i, x_j) + \lambda^2\delta_{ij}
\end{equation}
where $\delta_{ij}$ is the Kronecker delta. 

\begin{figure}[!t]
\centering
\parbox{\linewidth}{
\vspace{-0.01cm}
\hspace{-3.5cm}(a)
\vspace{0.01cm}\\
\centering
\scalebox{0.7}{
\begin{quantikz} 
    \lstick{$|0\rangle$} 
    & \gate{R_y(\theta_0)} 
    & \gate{R_z(x)} 
    \gategroup[2,steps=2,
    style={dashed,rounded corners,fill=blue!20},
    background,
    label style={label position=below,anchor=north,yshift=-0.2cm}]
    {\sc \small Layer repeated L times} 
    & \ctrl{1} 
    & \gate{R_y(\theta_3)} 
    & \\
    \lstick{$|0\rangle$} 
    & \gate{R_y(\theta_1)} 
    & \gate{R_z(x)} 
    & \gate{R_z(\theta_2)} 
    & \gate{R_y(\theta_4)} 
    &
\end{quantikz}
}
}

\vspace{0.35cm}

\parbox{\linewidth}{
\vspace{-0.2cm}
\hspace{-3.5cm}(b)
\vspace{0.2cm}\\
\centering
\scalebox{0.65}{
\begin{quantikz} 
    \lstick{$|0\rangle$} 
    & \gate{R_y(\theta_0)} 
    & \gate{R_x(\theta_0\cos^{-1}(x))} 
    \gategroup[2,steps=3,
    style={dashed,rounded corners,fill=blue!20},
    background,
    label style={label position=below,anchor=north,yshift=-0.2cm}]
    {\sc \small Layer repeated L times} 
    & \ctrl{1} 
    & \gate{R_z(\theta_3)} 
    & \gate{R_y(\theta_4)} 
    & \\
    \lstick{$|0\rangle$} 
    & \gate{R_y(\theta_1)} 
    & \gate{R_x(\theta_1\cos^{-1}(x))} 
    & \gate{R_z(\theta_2)} 
    & \ctrl{-1} 
    & \gate{R_y(\theta_5)} 
    &
\end{quantikz}
}
}
\caption{Quantum feature maps considered in this work: (a) Simple feature map with repeated linear entanglement, and (b) Chebyshev feature map with repeated circular entanglement. Redrawn from~\cite{rapp2024quantum}.}
\label{fig:feature-maps}
\end{figure}
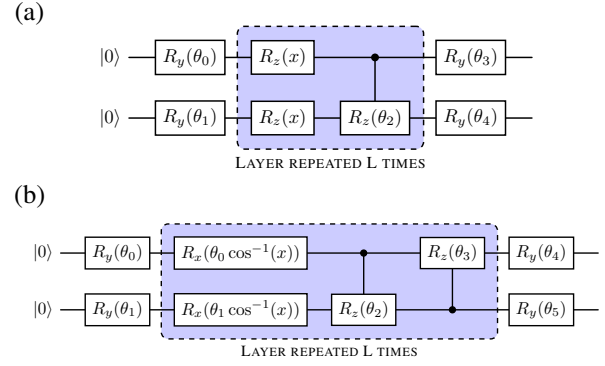

\subsection{Quantum Kernels}

The covariance function in the surrogate can be replaced by a quantum kernel while preserving the probabilistic structure required by the acquisition function for sequential sampling of evaluation points. A quantum kernel is constructed using a quantum feature map that encodes classical inputs $x$ to quantum states $|\phi(x) \rangle$ using a parameterized circuit. In this work, we will use the fidelity kernel, defined as $\Sigma(x, x') = |\langle \phi(x)|\phi(x') \rangle|^2$.

\section{Experimental Setup}

We empirically evaluate the active learning workflow with GPR to model the one-dimensional function $f(x) = x \sin(x)$, where $x$ is sampled uniformly from the interval $[0, 2\pi]$. The surrogate is trained with 30 data points and the active learning workflow is run for $10$ iterations to simulate a limited evaluation budget. We evaluate the mean squared error (MSE) on a test set of $m=20$ data points as our measure of performance. We explore how regularization of hyperparameters impacts performance and consider quantum kernels constructed from the feature maps shown in Fig.~\ref{fig:feature-maps}, which vary in the complexity of the entanglement they generate at short depth.

\begin{figure}[!t]
    \centering
    \includegraphics[
    width=0.4\textwidth,
    trim=2.25cm 3.35cm 2cm 5.5cm
    ]{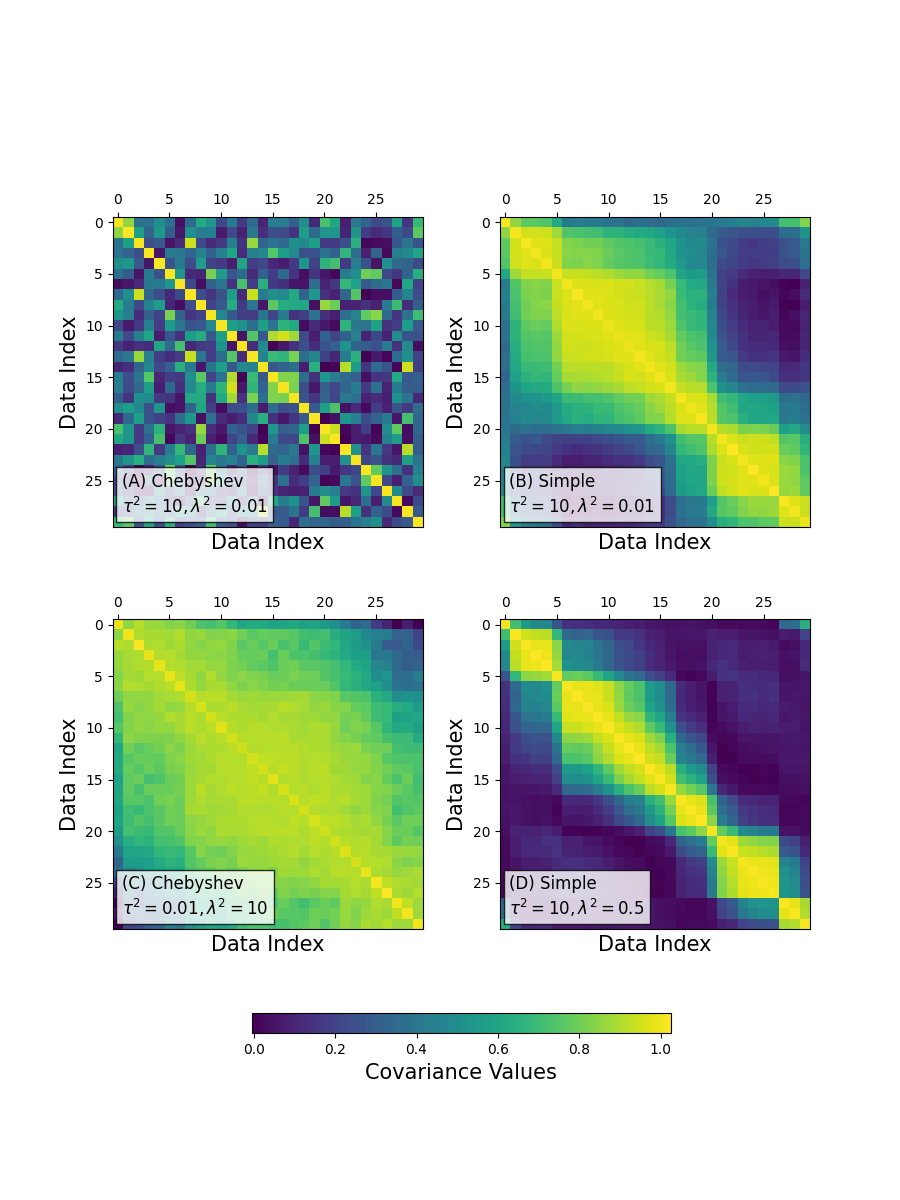}
    \caption{Normalized covariance matrices obtained after training the QGPR surrogate. (A) Displays overfitting in the $\tau^2 > \lambda^2$ regime. (B) Shows limited expressivity. (C) Shows underfitting with $\tau^2 < \lambda^2$. (D) Displays balanced performance.}
    \label{fig:kernel_matrices}
\end{figure}

We refer to the circuits in Fig.~\ref{fig:feature-maps}(a) and Fig.~\ref{fig:feature-maps}(b) as Simple and Chebyshev feature maps or kernels, respectively~\cite{rapp2024quantum,kyriienko2021solving}. Each kernel is constructed with $N =2$ qubits and $L=2$ layers, and Simple also with $N =4$ qubits. When using the Chebyshev feature map, the input features are normalized to the range $[-1, 1]$ due to the $\cos^{-1}$ preprocessing, while this normalization is not required for the Simple feature map. 

We also consider two types of kernel training: single-shot and iterative. In the single-shot kernel training method, the quantum kernel is trained to obtain optimal parameter values only once at the beginning of each active learning workflow and remains fixed for subsequent iterations. In the iterative kernel training method, the quantum kernel is trained at each iteration. We utilize Qiskit~\cite{qiskit2024} for kernel training and develop custom Python classes to interface with the existing GaussianProcessRegressor class in scikit-learn~\cite{scikit-learn}. All experiments are executed on Qiskit's AerSimulator. Kernel training is performed using the SPSA optimizer~\cite{spall1998overview} which minimizes Negative Log Marginal Likelihood (NLML) loss to obtain optimal kernel parameters. 

\section{Results and Discussion}

We run a total of 5 experiments across the various combinations of Simple and Chebyshev kernels, implemented according to a single-shot or iterative training method. Each experiment is run over 25 possible combinations of $(\tau^2, \lambda^2)$, each fixed with a value in $\{0.01, 0.1, 0.5, 1, 10\}$. Throughout the experiments, Simple and Chebyshev feature maps yield similar best-case performance, measured by MSE. For each feature map, the iterative and single-shot training methods perform comparably, with no qualitative differences observed. We only report the MSE values obtained from the iterative training for brevity in Tables~\ref{table:trainable_simple_n2}, \ref{table:trainable_simple_n4}, and \ref{table:trainable_chebyshev_n2}. The omitted results do not affect the conclusions of the paper. Overall, we identify four different performance regimes as a function of the hyperparameters, with qualitatively distinct kernel structures, Fig.~\ref{fig:kernel_matrices}, and surrogate models, Fig.~\ref{fig:qgpr_performance}.

\begin{table}[!t]
\centering
\footnotesize
\resizebox{0.38\textwidth}{!}{
\begin{tabular}{c|ccccc}
\diagbox{$\tau^2$}{$\lambda^2$}
    & 0.01 & 0.1 & 0.5 & 1 & 10 \\ \hline
0.01 &
\cellcolor{CellOrange!30} 0.40 &
\cellcolor{teal!30} 5.25 &
\cellcolor{teal!30} 4.08 &
\cellcolor{teal!30} 5.55 &
\cellcolor{teal!30} 5.79 \\

0.1 &
\cellcolor{CellOrange!30} 0.06 &
\cellcolor{CellOrange!30} 0.12 &
\cellcolor{CellOrange!30} 0.79 &
\cellcolor{CellOrange!30} 1.70 &
\cellcolor{teal!30} 5.82 \\

0.5 &
\cellcolor{CellOrange!30} 0.3 &
\cellcolor{CellOrange!30} 0.02 &
\cellcolor{CellOrange!30} 0.10 &
\cellcolor{CellOrange!30} 0.23 &
\cellcolor{teal!30} 5.22 \\

1 &
\cellcolor{magenta!30} 4.1 &
\cellcolor{CellOrange!30} 0.05 &
\cellcolor{CellOrange!30} 0.06 &
\cellcolor{CellOrange!30} 0.09 &
\cellcolor{CellOrange!30} 3.28 \\

10 &
\cellcolor{magenta!30} 198.01 &
\cellcolor{CellOrange!30} 0.23 &
\cellcolor{CellOrange!30} 0.05 &
\cellcolor{CellOrange!30} 0.03 &
\cellcolor{CellOrange!30} 0.42 \\
\end{tabular}
}
\vspace{5pt}
\caption{MSE values from Iterative Training with Simple Kernel (N=2).}
\label{table:trainable_simple_n2}
\end{table}

\begin{table}[!t]
\centering
\footnotesize
\resizebox{0.38\textwidth}{!}{
\begin{tabular}{c|ccccc}
\diagbox{$\tau^2$}{$\lambda^2$}
    & 0.01 & 0.1 & 0.5 & 1 & 10 \\ \hline
0.01 &
\cellcolor{CellOrange!30} 0.20 &
\cellcolor{CellOrange!30} 1.99 &
\cellcolor{CellOrange!30} 4.17 &
\cellcolor{CellOrange!30} 5.24 &
\cellcolor{teal!30} 5.97 \\

0.1 &
\cellcolor{CellOrange!30} 0.003 &
\cellcolor{CellOrange!30} 0.09 &
\cellcolor{CellOrange!30} 0.93 &
\cellcolor{CellOrange!30} 1.97 &
\cellcolor{teal!30} 6.04 \\

0.5 &
\cellcolor{CellOrange!30} 0.04 &
\cellcolor{CellOrange!30} 0.01 &
\cellcolor{CellOrange!30} 0.11 &
\cellcolor{CellOrange!30} 0.26 &
\cellcolor{CellOrange!30} 3.70 \\

1 &
\cellcolor{CellOrange!30} 0.02 &
\cellcolor{CellOrange!30} 0.02 &
\cellcolor{CellOrange!30} 0.03 &
\cellcolor{CellOrange!30} 0.09 &
\cellcolor{CellOrange!30} 1.98 \\

10 &
\cellcolor{CellOrange!30} 2.09 &
\cellcolor{CellOrange!30} 0.03 &
\cellcolor{CellOrange!30} 0.01 &
\cellcolor{CellOrange!30} 0.01 &
\cellcolor{CellOrange!30} 0.28 \\
\end{tabular}
}
\vspace{5pt}
\caption{MSE values from Iterative Training with Simple Kernel (N=4).}
\label{table:trainable_simple_n4}
\end{table}

\begin{table}[!t]
\centering
\footnotesize
\resizebox{0.38\textwidth}{!}{
\begin{tabular}{c|ccccc}
\diagbox{$\tau^2$}{$\lambda^2$}
    & 0.01 & 0.1 & 0.5 & 1 & 10 \\ \hline
0.01 &
\cellcolor{Periwinkle!30} 7.65 &
\cellcolor{Periwinkle!30} 4.51 &
\cellcolor{teal!30} 5.96 &
\cellcolor{teal!30} 6.24 &
\cellcolor{teal!30} 6.75 \\

0.1 &
\cellcolor{Periwinkle!30} 8.64 &
\cellcolor{Periwinkle!30} 5.32 &
\cellcolor{CellOrange!30} 2.06 &
\cellcolor{teal!30} 4.5 &
\cellcolor{teal!30} 6.81 \\

0.5 &
\cellcolor{Periwinkle!30} 34.73 &
\cellcolor{Periwinkle!30} 5.10 &
\cellcolor{CellOrange!30} 0.11 &
\cellcolor{CellOrange!30} 1.43 &
\cellcolor{teal!30} 6.41 \\

1 &
\cellcolor{Periwinkle!30} 155.09 &
\cellcolor{Periwinkle!30} 9.05 &
\cellcolor{CellOrange!30} 0.05 &
\cellcolor{CellOrange!30} 0.33 &
\cellcolor{teal!30} 5.84 \\

10 &
\cellcolor{Periwinkle!30} 10466.70 &
\cellcolor{Periwinkle!30} 113.00 &
\cellcolor{Periwinkle!30} 14.76 &
\cellcolor{CellOrange!30} 0.01 &
\cellcolor{teal!30} 3.69 \\
\end{tabular}
}
\vspace{5pt}
\caption{MSE values from Iterative Training with Chebyshev Kernel (N=2).}
\label{table:trainable_chebyshev_n2}
\end{table}

\subsection{Overfitting}
When $\tau^2 > \lambda^2$, the GPR overfitted the training data due to the high weighting of the covariance matrix. This behavior is observed in the Chebyshev kernel, as highlighted by the purple MSE values in Table~\ref{table:trainable_chebyshev_n2}. In this case, the covariance matrix, Fig.~\ref{fig:kernel_matrices}(A), is diagonal and the surrogate has no global structure, as seen by the purple dots in Fig.~\ref{fig:qgpr_performance}. While this is a hallmark of overfitting, the model is not only not generalizable, but it has even failed to accurately predict the training data.

\subsection{Limited Expressivity}
The overfitting observed for the Chebyshev kernel, when $\tau^2 > \lambda^2$, is only possible because the feature map is expressive enough to capture distinctions within the training data. When the feature map is insufficiently expressive, as in the case of Simple kernel with $N=2$ qubits, we observe that the kernel groups the training points into highly correlated regions, Fig.~\ref{fig:kernel_matrices}(B), and the surrogate shows little structure, as seen by the pink dots in Fig.~\ref{fig:qgpr_performance}. While this kernel may appear to encode useful correlations, it results in poor performance as seen by the pink MSE values in Table~\ref{table:trainable_simple_n2}. Increasing the width of the feature map to $N=4$ qubits improves the expressivity of the feature map so that this is no longer an issue, as seen by the absence of pink colored MSE values in Table~\ref{table:trainable_simple_n4}.

\subsection{Underfitting}
On the other hand, when $\tau^2 < \lambda^2$, synthetic noise dominates kernel training, resulting in poor performance indicated by the large teal colored MSE values in all tables. A sample covariance matrix from this region is nearly an all-unit matrix, Fig.~\ref{fig:kernel_matrices}(C), indicating that the model has not learned any relationship within the training data, and the surrogate locks to a global mean value seen by the string of teal dots in Fig.~\ref{fig:qgpr_performance}. We only show the quantum covariance matrix to highlight the impact of the noise variance hyperparameter on the model training. The fully regularized kernel with the component $\lambda^2\delta_{ij}$ is a nearly diagonal matrix as $\lambda$ dominates $\tau$.

\subsection{Balanced Performance}
\begin{figure}[!t]
    \centering
    \includegraphics[
    width=0.4\textwidth,
    trim=1.5cm 0.5cm 0.5cm 1cm
    ]{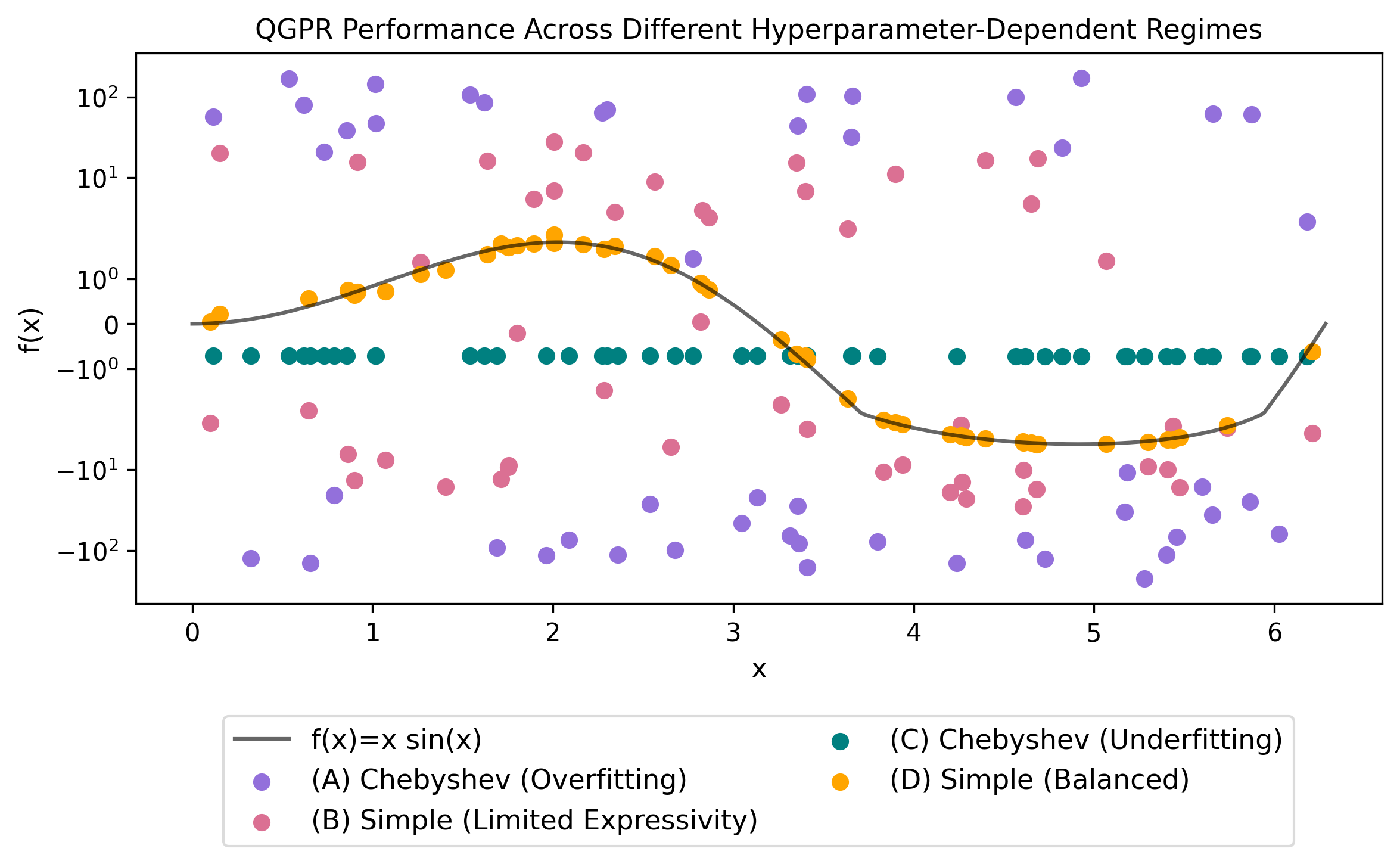}
    \caption{QGPR performance across the four hyperparameter-dependent regimes. The colored dots depict the QGPR mean prediction for each model.}
    \label{fig:qgpr_performance}
\end{figure}

We observe good performance in intermediate regions of the regularization hyperparameters, especially for larger values of $\tau^2$ with moderate $\lambda^2$, indicated by the orange MSE values in all tables. The covariance matrix for the lowest MSE obtained is shown in Fig.~\ref{fig:kernel_matrices}(D). The kernel encodes the local correlations between nearby input points consistent with the simple black-box function, and the surrogate is an excellent approximation, as seen by the orange dots along the original function plot in Fig.~\ref{fig:qgpr_performance}.

In our simulations, low MSE values correspond to kernels that present balanced performance in modeling a black-box function. However, high MSE values are observed in the underfitting, overfitting, and limited expressivity regimes. The MSE values alone cannot distinguish this behavior and the cause of poor performance. The structure of the learned kernel, Fig.~\ref{fig:kernel_matrices}, must also be examined. In comparison, only underfitting was observed for a classical RBF kernel, while it yielded low MSE and consistent covariance matrix once $\tau^2>\sigma^2$.

\section{Conclusion}
In this work, we explored the application of quantum kernels to GPR in an active learning framework for modeling black-box functions. Studying a regression problem is important since they are prone to overfitting and have the potential for improved performance with regularization. We focused on the impact of regularization through hyperparameter tuning and on how it balances expressivity and overfitting. While kernel methods alone are sensitive to the relative scale of the regularization hyperparameters, active learning depends on the individual values of $\tau^2$ and $\sigma^2$. We found that overfitting can manifest itself as distinct kernel correlation patterns, but good performance can be found with appropriate regularization.

We investigated small and shallow quantum feature maps to simplify the interpretability of the results, and future work should explore how kernel regularization can be deployed for more complex black-box functions and with larger quantum feature maps. While regularization scales naturally with system size, care must be taken to find the appropriate hyperparameters by balancing against the average magnitude of covariance values. The interpretation of regularization as ``noise'', and whether this can be implemented directly within a quantum model, e.g.~in quantum reservoir computing~\cite{GoviaRC}, is also an interesting area of future study.

\section*{Acknowledgment}
This work was supported by the FABrIC project managed by CMC Microsystems and funded by the Government of Canada. 

\bibliographystyle{IEEEtran}
\bibliography{references}

\end{document}